\documentclass{aa}  

\usepackage{graphicx}
\usepackage{txfonts}
\usepackage{siunitx}
\usepackage{natbib}
\usepackage{amsmath}
\usepackage{subfigure}
\usepackage{hyperref}
\hypersetup{
    colorlinks=true,
    citecolor=blue,
    linkcolor=blue,
    urlcolor=blue,
}

\usepackage{orcidlink}
\newcommand{\orcid}[1]{\href{https://orcid.org/#1}
{{\includegraphics[height=8pt]{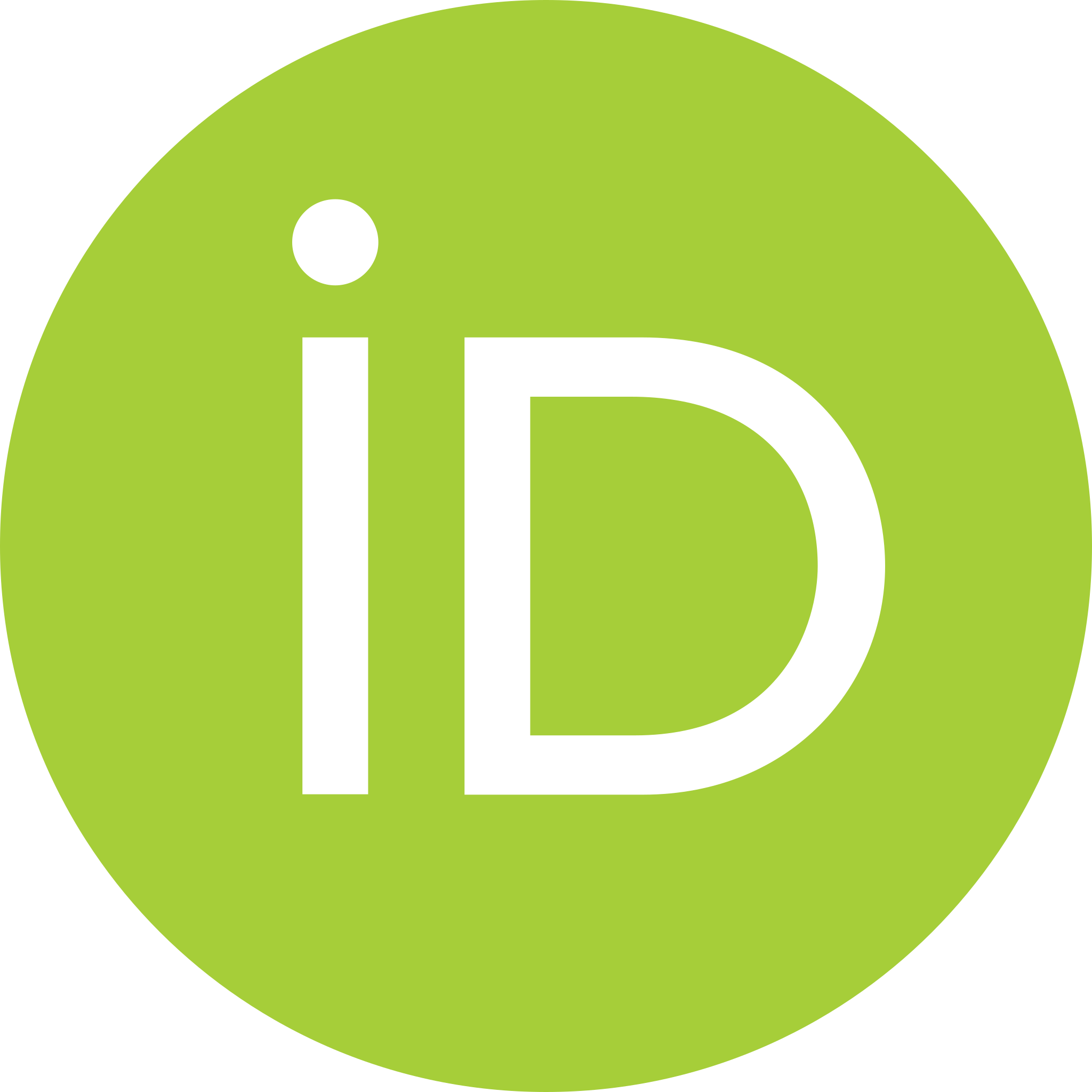}}}}

\begin{document}

   \title{Recovered SN Ia rate from simulated LSST images}
            
   \author{V. Petrecca\thanks{vincenzo.petrecca@inaf.it}\inst{1,2}\orcid{0000-0002-3078-856X}
          \and
          M. T. Botticella\inst{1}\orcid{0000-0002-3938-692X}
          \and
          E. Cappellaro\inst{3}\orcid{0000-0001-5008-8619}
          \and 
          L. Greggio\inst{3}\orcid{0000-0003-2634-4875}
          \and
          B. O. Sánchez\inst{4}\orcid{0000-0002-8687-0669}
          \and
          A. Möller\inst{5}\orcid{0000-0001-8211-8608}
          \and
          M. Sako\inst{6}\orcid{0000-0003-2764-7093}
          \and
          M. L. Graham\inst{7}\orcid{0000-0002-9154-3136}
          \and
          M. Paolillo\inst{2,1,8}\orcid{0000-0003-4210-7693}
          \and
          F. Bianco\inst{9,10}\orcid{0000-0003-1953-8727}
          \and
          the LSST Dark Energy Science Collaboration
          }

   \institute{INAF - Osservatorio Astronomico di Capodimonte, Salita Moiariello 16, 80131 Napoli, Italy
        \and 
        Department of Physics, University of Napoli “Federico II”, via Cinthia 9, 80126 Napoli, Italy
        \and
        INAF – Osservatorio Astronomico di Padova, Vicolo dell’Osservatorio 5, Padova 35122, Italy
        \and
        Aix Marseille Univ, CNRS/IN2P3, CPPM, Marseille, France
        \and
        Centre for Astrophysics and Supercomputing, Swinburne University of Technology, Mail Number H29, PO Box 218, 31122 Hawthorn, VIC, Australia
        \and
        Department of Physics and Astronomy, University of Pennsylvania, Philadelphia, PA 19104, USA
        \and
        DIRAC Institute, Department of Astronomy, University of Washington, 3910 15th Avenue NE, Seattle, WA 98195, USA
        \and
        INFN — Sezione di Napoli, via Cinthia 9, 80126, Napoli, Italy
        \and
        University of Delaware, 210 South College Ave., Newark, DE  19716, USA
        \and
        Vera C. Rubin Observatory, Tucson, AZ 85719, USA
        }

   \date{Received 19 December 2023 / Accepted 23 February 2024}

 
  \abstract
   {}
   {The Vera C. Rubin Observatory's Legacy Survey of Space and Time (LSST) will revolutionize Time Domain Astronomy by detecting millions of different transients. In particular, it is expected to increment the number of type Ia supernovae (SN Ia) of a factor of 100 compared to existing samples up to redshift $\sim 1.2$. Such a high number of events will dramatically reduce statistical uncertainties in the analysis of SN Ia properties and rates. However, the impact of all other sources of uncertainty on the measurement of the SN Ia rate must still be evaluated. The comprehension and reduction of such uncertainties will be fundamental both for cosmology and stellar evolution studies, as measuring the SN Ia rate can put constraints on the evolutionary scenarios of different SN Ia progenitors.}
   {We use simulated data from the Dark Energy Science Collaboration (DESC) Data Challenge 2 (DC2) and LSST Data Preview 0 to measure the SN Ia rate on a 15 $deg^2$ region of the Wide-Fast-Deep area. We select a sample of SN candidates detected on difference images, associate them to the host galaxy with a specially developed algorithm, and retrieve their photometric redshifts. Then, we test different light curves classification methods, with and without redshift priors (albeit ignoring contamination from other transients, as DC2 contains only SN Ia). We discuss how the distribution in redshift measured for the SN candidates changes according to the selected host galaxy and redshift estimate.}
   {We measure the SN Ia rate analyzing the impact of uncertainties due to photometric redshift, host galaxy association and classification on the distribution in redshift of the starting sample. We found a 17\% average lost fraction of SN Ia with respect to the simulated sample. As 10\% of the bias is due to the uncertainty on the photometric redshift alone (which also affects classification when used as a prior), it results to be the major source of uncertainty. We discuss possible reduction of the errors in the measurement of the SN Ia rate, including synergies with other surveys, which may help using the rate to discriminate different progenitor models.}
   {}

   \keywords{Stars: supernovae --
                Galaxies: stellar content: surveys
               }

   \maketitle
%

\section{Introduction}\label{sec:intro}
Type Ia supernovae (SN Ia) are violent explosions characterized by a peak in luminosity correlated to the duration of the event, which makes them standardizable candles (\citealt{1993ApJ...413L.105P}; \citealt{1999ApJ...525..209T}) and fundamental cosmological probes (\citealt{1998AJ....116.1009R}; \citealt{1999ApJ...517..565P}). There is general consensus that SN Ia are the result of a thermonuclear explosion of a carbon-oxygen White Dwarf (WD) with two possible progenitor channels: a WD accreting mass from a non-degenerate star (single degenerate scenario SD; \citealt{1973ApJ...186.1007W}) or two WD spiraling together and eventually merging (double degenerate scenario DD; \citealt{1984ApJ...277..355W}; \citealt{1984ApJS...54..335I}). However, the exact nature of their progenitors and the details of the explosion mechanism are not yet clear (see \citealt{2018PhR...736....1L} for a recent review). Direct observations of both pre- and post- explosion images do not provide unambiguous evidence for either SN Ia progenitor systems (e.g., \citealt{2014ApJ...790....3K}; \citealt{2019MNRAS.484L..79G}), or surviving companions (e.g., \citealt{2012Natur.481..164S}; \citealt{2019A&A...623A..34K}). Similarly, detailed spectral analyses of SN Ia and their remnants are not able to clearly identify the companion star in the binary system (e.g., \citealt{2007ApJ...662..472B}; \citealt{2012ApJ...752..101F}; \citealt{2016A&A...588A..84D}). The diversity of the SN Ia light curves and their correlation with the host galaxy properties are also not able to exclude any progenitor scenario, hence both of them might be at play.

An alternative way of putting constraints on the SN Ia progenitors system is the analysis of the delay times distribution (DTD), which is the time between the formation of the binary system and the SN Ia explosion. Different progenitor scenarios imply a different DTD from population synthesis models (\citealt{2012NewAR..56..122W}). As the SN Ia rate results from the convolution of the host galaxy star formation rate (SFR) and the DTD (\citealt{2005A&A...441.1055G}; \citealt{2010MNRAS.406...22G}), measuring the SN Ia rate and the SFR for a sample of galaxies is thus an empirical way of testing the DTD from different progenitor models (\citealt{2012MNRAS.426.3282M}; \citealt{2019A&A...625A.113G}; \citealt{2020ApJ...890..140S}; \citealt{2021MNRAS.506.3330W}). More into detail, the SN Ia rate at a time \textit{t}, $r_{SN Ia}(t)$ can be expressed as:
\begin{equation}
    r_{SN Ia}(t) = k_{Ia} \int_0^t \psi(t-t_D)f_{Ia}(t_D)dt_D,
    \label{eq:rate_models}
\end{equation}
where $k_{Ia}$ is the total number of SN Ia provided by a stellar population of unitary mass, $\psi$ is the SFR, $f_{Ia}$ is the DTD, and $t_D$ is the delay time. Knowing the average cosmic star formation history (SFH) and the DTD for each progenitor system model, it is also possible to calculate the expected volumetric rate of SN Ia as a function of redshift and compare it with the observed rate. 

\begin{figure}
    \includegraphics[width=1.0\columnwidth]{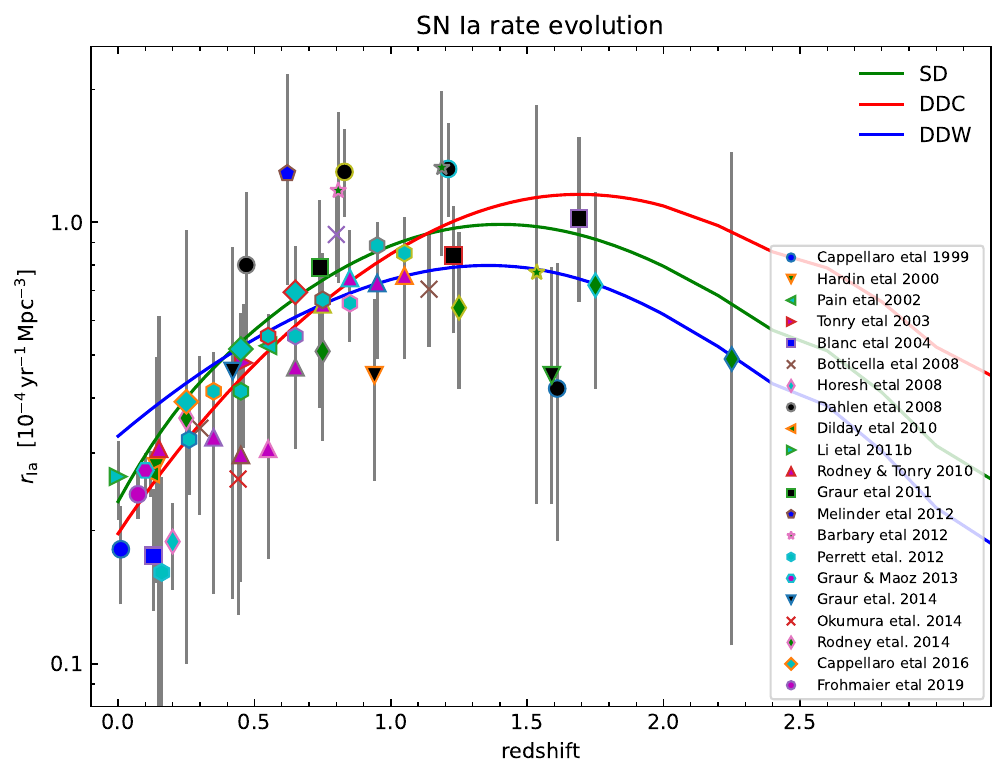}
    \caption{Observed SN Ia rate as a function of redshift for different surveys, along with rate predictions for progenitor models from \cite{2005A&A...441.1055G}: Single Degenerate (SD), Double Degenerate Close (DDC), Double Degenerate Wide (DDW).}
    \label{fig:rate_enrico}
\end{figure}

Figure \ref{fig:rate_enrico} shows the measurements of the SN Ia rate as a function of redshift from different surveys, along with rate predictions for DTD progenitor models from \cite{2005A&A...441.1055G}, adopting the estimates of cosmic SFH from \cite{2017ApJ...840...39M}. Theoretical rates have been normalized using the same $k_{Ia} = 0.8 \times 10^{-3}M_{\odot}^{-1}$ for all the models. The significant scatter between rate measurements from different surveys, as well as present day statistical and systematic uncertainties on each single measurement, do not allow us to distinguish between different progenitor models. Moreover, the theoretical predictions run quite close in the intermediate range $0.2 \lessapprox z \lessapprox 1.0$, making it difficult to discriminate among the different options. \cite{2019A&A...625A.113G} show that measuring the SN Ia rate as a function of host galaxy intrinsic colors or specific SFR is more efficient in separating the predictions of different models, but data are only limited to the local Universe. Upcoming surveys, such as the Vera C. Rubin Observatory's Legacy Survey of Space and Time (LSST;\footnote{\hyperlink{http://www.lsst.org}{lsst.org}} \citealt{2019ApJ...873..111I}), may completely change the scenario. Indeed LSST will detect an enormous number of events in galaxies with a large range of different properties, strongly improving on both statistical and systematic uncertainties. Besides, having data from the same survey, with known and homogeneous properties, will also reduce the scatter between rate measurements in different redshift and intrinsic color bins, compared to those coming from the combination of multiple surveys. While a dramatic reduction of statistical uncertainties will be easily attained by LSST, the actual effect of all other possible sources of uncertainty deserves a detailed analysis. Some of the uncertainties are not directly related to the survey and depend on the adopted software/criterion for the analysis (e.g., photometric redshift, transient classification) and the wealth of ancillary data (e.g., spectroscopic information). 

In this work, we will assess the impact of uncertainties on the SN Ia rate using a simulation of the first 5 years of LSST. As the simulation does not include other SN types nor any correlation between the transients and the host galaxy properties, we only focus on evaluating the impact of uncertainties on the SN Ia rate as a function of redshift due to i) host galaxy association, ii) photometric redshift,  iii) and light curves classification, using a sample of SN Ia detected on difference images (see Sect. \ref{sec:data}). These sources of uncertainty, affecting the choice of the SN Ia sample and their host galaxies, also reflect on rate measurements as a function of specific SFR or intrinsic galaxy colors from spectral energy distribution (SED) fitting. Furthermore, the sources of error studied in this work are common in all transients studies, and understanding how they propagate in a final statistical analysis would be important for most of Time Domain science with LSST. Measuring the SN rate and shedding light on the evolutionary channels for SNe are also goals outlined in both Dark Energy Science Collaboration\footnote{\hyperlink{https://lsstdesc.org/}{lsstdesc.org}} (DESC) and Transient and Variable Stars Science Collaboration\footnote{\hyperlink{https://lsst-tvssc.github.io/}{lsst-tvssc.github.io}} (TVSSC) roadmaps\footnote{DESC roadmap: \hyperlink{https://lsstdesc.org/assets/pdf/docs/DESC_SRM_latest.pdf}{lsstdesc.org/assets/pdf/docs/DESC\_SRM\_latest.pdf}} (\citealt{tvs_roadmap}). The comprehension of all possible biases affecting the final measurement is thus a crucial point before the beginning of the survey. A more thorough investigation on the SN Ia rate from LSST as a tool to discriminate between progenitor models is left to subsequent analysis. 

This paper is organized as follows. Section \ref{sec:LSST} provides a short summary of the LSST strategy and design. In Sect. \ref{sec:data} we briefly describe the simulation and the SN sample used in our analysis. In Sect. \ref{sec:host}, we describe the procedure of association of transients to the host galaxies, and in Sect. \ref{sec:photoz} the photometric redshifts of the hosts. Section \ref{Sec:class} is dedicated to the classification of light curves and the comparison of different approaches. Finally, we present the final measurement of the SN Ia rate in Sect. \ref{sec:rate}, comparing the results with the input of the simulation and providing quantitative estimates of different sources of uncertainty. In Sect.\ref{sec:concl}, we summarize our results and discuss future perspectives.


\section{Rubin Observatory's LSST}\label{sec:LSST}
The LSST, expected to start in 2025\footnote{\hyperlink{https://www.lsst.org/about/project-status}{lsst.org/about/project-status}}, will revolutionize Time Domain astronomy by imaging the entire southern sky every few nights for 10 years (\citealt{2019ApJ...873..111I}). It will be executed with the 8.4 m (6.7 m effective) Simonyi Telescope and a 3.2 gigapixel camera yielding a $9.6\ deg^2$ field of view. The instrument is equipped with 6 filters \textit{ugrizy} and is expected to have a $5\sigma$ \textit{r}-band magnitude depth of $\sim 24.5$ in a single 30 seconds visit and $\sim 27.8$ in the full stacked data. The survey design will enable to cover a wide range of science goals (\citealt{2009arXiv0912.0201L}). The main ones are i) exploring the transient and variable sky, ii) studying the dynamics of solar system objects, iii) probing dark energy and dark matter, iv) and mapping the Milky Way. Fulfilling most of these science goals requires scanning the sky deep, wide and fast. Albeit the exact survey strategy is not yet defined, about 90\% of the observing time will be devoted to the baseline wide-fast-deep (WFD) survey mode. The remaining 10\% will be used to obtain improved coverage and cadence for specific regions, called deep drilling fields (DDF). 

The exploration of the changing universe will be boosted by the implementation of difference image analysis (DIA; \citealt{1998ApJ...503..325A}) on the entire dataset. This technique consists of producing deep co-added images (templates) to be subtracted from each science observation in the same region of sky. Before the subtraction, the template is re-sampled to the pixel coordinate system of the new image and is convolved with a kernel matching their Point Spread Functions (PSFs). Variable sources are considered to be detected if they have a signal-to-noise ratio (S/N) greater than a threshold value (i.e., five for Rubin data products) on the resulting difference image. For each detected transient, LSST will issue an \textit{alert} within 60 seconds of the end of the visit (defined as the end of image readout from the camera) to enable immediate follow-up observations with other facilities. 

LSST is expected to process $\sim 10^5$ transient detections per night\footnote{\hyperlink{https://www.lsst.org/scientists/keynumber}{lsst.org/scientists/keynumber}} (\citealt{2019ApJ...873..111I}; \citealt{DMTN-102}). This includes SN Ia for which we expect an increment of a factor of 100 compared to existing samples up to redshift z $\sim 1.2$ (e.g., \citealt{2014A&A...568A..22B}; \citealt{2018PASP..130f4002S}; \citealt{2019ApJ...881...19J}). Such a high number of events will dramatically reduce statistical uncertainties in the analysis of SN Ia properties and rates, which will be important both for cosmology and stellar evolution studies. The analysis of SNe is then expected to be only limited by the adopted observing strategy (affecting the sampling of light curves and the resulting classification of transients) and different sources of uncertainty, namely detection efficiency on difference images, photometric calibrations, artifact contamination levels, reliability of photometric redshifts of the host galaxies (e.g., precision, accuracy and fraction of catastrophic errors), efficiency of host galaxy association and photometric classification algorithms. Many of these aspects have been thoroughly looked into by recent works: \cite{2022ApJ...934...96S} measured detection efficiency, magnitude limits and photometric biases; \cite{2022ApJS..258...23A} and \cite{2018AJ....155....1G} focused on the impact of different observing cadences on the classification of SNe and the reliability of photometric redshifts; \cite{2020MNRAS.499.1587S} compared different algorithms for photometric redshifts; the Photometric LSST Astronomical Time-series Classification Challenge (\texttt{PLAsTiCC}; \citealt{2018arXiv181000001T}; \citealt{2019PASP..131i4501K}) and its extension \texttt{ELAsTiCC}\footnote{\hyperlink{https://portal.nersc.gov/cfs/lsst/DESC_TD_PUBLIC/ELASTICC/}{portal.nersc.gov/cfs/lsst/DESC\_TD\_PUBLIC/ELASTICC}} enable testing of different classification methods on light curves. However, all these works are oriented towards obtaining a pure sample of SN Ia for cosmology, while a study on the combination of multiple selection effects and sources of uncertainty on the different science case of SN rate measurement (which requires having a complete sample of SNe, even if with lower purity) is still lacking. Our work acts as a complement of the previous analyses and aims at determining the impact of host galaxy association, photometric redshift and transients classification on the measurement of the SN Ia volumetric rate using a simulation of the first 5 years of LSST. We discuss all these effects one by one, to understand what should be improved to obtain an accurate evaluation of the SN Ia rate up to $z \sim 1$ with LSST.


\section{The LSST DESC DC2 Universe}\label{sec:data}
To build software pipelines ready to analyze the LSST data products, DESC produced a $300\ deg^2$ simulation of the first 5 years of survey as part of the Data Challenge 2 (DC2; \citealt{DC2}). The simulation includes LSST-like images in all six optical bands \textit{ugrizy}, processed with the LSST Science Pipelines\footnote{\hyperlink{https://pipelines.lsst.io/}{pipelines.lsst.io}} (v.19.0.0). It is based on the \textit{Outer Rim N}-body simulation (\citealt{2019ApJS..245...16H}), while the observing cadence is determined by the \textit{minion\_1016}\footnote{\hyperlink{https://docushare.lsst.org/docushare/dsweb/View/Collection-4604}{https://docushare.lsst.org/docushare/dsweb/View/Collection-4604}} strategy for the WFD survey with an average cadence of $\sim 3$ days in any filter. A smaller $1\ deg^2$ DDF with more frequent observations (up to one per day) is also included.

Simulated sources comprise stars, galaxies, variable stars and SN Ia (but no other SN types). SN Ia light curves are simulated starting from the rest-frame Spectral Energy Distribution (SED) of the SALT2-Extended model (\citealt{2010A&A...523A...7G}; \citealt{2018PASP..130k4504P}), which uses 5 parameters: redshift ($z$), time at peak brightness ($t_0$), amplitude ($x_0$), stretch ($x_1$), and color ($c$). The redshift distribution of SN Ia follows the volumetric rate $r_v(z)=2.5\times10^{-5}(1+z)^{1.5}Mpc^{-3}yr^{-1}$ (\citealt{2008ApJ...682..262D}) up to $z=1.4$. SNe were assigned to galaxies with an occupation probability which scales with the galaxy mass. The SN position within the host traces the light in the galaxy sampled by the surface brightness profile of the disk and the bulge. SNe at $z>1.0$ are not associated to galaxies while at lower redshift, 10\% of SNe were randomly injected as host-less. Correlations between the SN type and the host galaxy properties are not included in the simulation.

The DC2 WFD region is also included in the LSST Data Preview 0 (DP0\footnote{\hyperlink{https://dp0-2.lsst.io/}{dp0-2.lsst.io}}), the first of three data previews serving as an early integration test of the LSST Science Pipelines and the Rubin Science Platform (RSP). A limited number of data rights holders has been granted access to the RSP, in order to start to familiarize with the Rubin Data Products\footnote{Data Products Definitions Document (DPDD; \hyperlink{https://ls.st/dpdd}{ls.st/dpdd})} using a series of publicly available tutorials\footnote{\hyperlink{https://github.com/rubin-dp0}{github.com/rubin-dp0}}. DP0 has been released in three parts: 
\begin{itemize}
    \item DP0.1 (released on June 30, 2021) is the DESC processing of the data, focusing on static sky;
    \item DP0.2 (released on June 30, 2022) is the DC2 simulation reprocessed by the Rubin staff using version 23 of the LSST Science Pipelines and, and includes DIA data products;
    \item DP0.3 (released on August 2, 2023), LSST-like catalogues of solar system objects generated by the Solar System Science Collaboration\footnote{\hyperlink{https://dp0-3.lsst.io/}{dp0-3.lsst.io}}.
\end{itemize}
For more official references for DP0, see \cite{RTN-001}, \cite{RTN-004}, and \cite{RTN-041}. All the galaxy catalogues used in this paper are extracted from DP0.2 and the codes used for the analysis have been successfully tested on the RSP, although the majority of the heavy tasks have been executed on DESC allocated space at NERSC\footnote{National Energy Research Scientific Computing Center; \hyperlink{https://www.nersc.gov/}{nersc.gov}}, which already contained all the necessary tables when this work started. 

The sample of SN Ia analyzed in this paper comes from the work by \cite{2022ApJ...934...96S}, which measured magnitude limits, detection efficiency, artifact contamination levels, and biases in the selection and photometry on a subset of $\sim 15\ deg^2$ of the DC2 WFD region. To perform the analysis, they tested the LSST DIA pipeline on the selected region, using the first year to produce their own templates and the remaining 4 years for detection. The smaller region of sky and the shorter survey length (5 years instead of the planned 10) do not affect the final results. 

\begin{figure}
    \includegraphics[width=1.0\columnwidth]{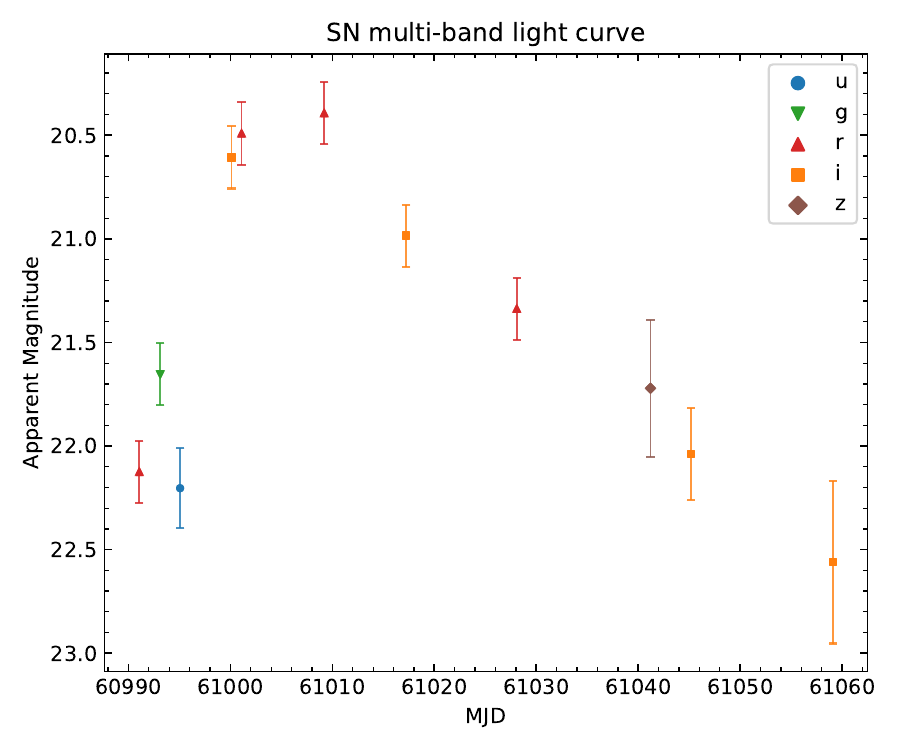}
    \caption{Example multi-band light curve of a SN Ia with redshift $z=0.16$. Different shapes and colors refer to different filters, as reported in the legend.}
    \label{fig:lc_example}
\end{figure}

There is a total of 5884 simulated SN Ia with $z\le 1.0$ in the processed area. This number takes already into account all the objects not detectable because of gaps between the observing seasons or located in sky regions with subtraction artifacts from template overlapping issues. The number of SNe with at least one detection in one filter is 2186. Yet, for our analysis, we selected only sources detected in more than 5 distinct nights, in order to sample $\sim 20$ days on the light curve and have a more reliable classification. In case of multiple visits in the same night, we took the average magnitude for each detection. An example of the recovered SN Ia light curve at $z = 0.16$ is shown in Fig. \ref{fig:lc_example}. The condition on the minimum number of detections returned 600 SNe. The large drop of SNe after the cut on the minimum number of detections is mainly due to the sub-optimal observing cadence of the adopted survey strategy. The \textit{minion\_1016} strategy adopted by DC2 is indeed quite old, albeit it was most recent at the times of the simulation. Further analyses to improve the observing strategy are still ongoing, and the final cadence may improve the number of recovered SNe. \cite{2022ApJS..259...58L} presented various metrics developed by DESC to analyze the cadences, highlighting the importance of low inter-night gaps in the redder filters for the selection of SNe. Possible improvements might also derive from a rolling cadence (\citealt{Alves+23}) or the higher coverage of the DDFs (\citealt{Gris+23}). 

The final sample of SN Ia analyzed in this work consists then of 600 sources (hereafter, SN Ia sample) for which we know the true host galaxy and the true simulated redshift (\textit{zspec}). Throughout the paper, we treat them as a sample of candidate SNe of unknown type, ignoring all known parameters from the simulation and using only the recovered photometric information. We first associate each SN to the host galaxy with a procedure described in Sect. \ref{sec:host}. Then we retrieve photometric redshifts for both the true (i.e., the simulation input) and the recovered host galaxies. Finally, we proceed with classification of light curves using the recovered redshifts as priors. Samples of SN Ia coming from each of these steps will produce a different distribution in redshift (see Sect. \ref{sec:photoz}), thus affecting the SN Ia volumetric rate. The real case scenario consists of a sample of SNe classified as SN Ia using the photometric redshift of the associated host galaxy as a prior. Such a scenario includes a combination of all the effects analyzed in this work. We compute the SN Ia rate for different samples, alternatively analyzing each specific source of uncertainty in order to analyze the contribution of each effect separately and determine which one impacts most on the final uncertainty. 


\section{Host galaxy association} \label{sec:host}
The association of SNe to the host galaxy is a key ingredient for the SN rate, but also for the SN photometric classification, as it provides an estimate of the SN redshift that can be used as prior in the classification (see Sect. \ref{Sec:class}). Indeed, as spectroscopic resources are limited, photometric redshifts of the host galaxies are an efficient way to obtain information on SN redshift. Moreover, the knowledge of the host galaxy and its properties enables the measurement of the SN rate as a function of the SFR, color, or the stellar mass of the galaxy (e.g., \citealt{2006ApJ...648..868S}; \citealp{2017A&A...598A..50B}), and provides constraints on the SN progenitors (\citealp{2021MNRAS.506.3330W}). 

We associate transients to galaxies using only information about the galaxy light profile and the angular separation from SNe. More complicated approaches could include any correlations between the SN type and the host galaxy properties, which are not included in this simulation. Recent reviews of host association algorithms, also using postage stamps of a field surrounding the transient, are reported in \cite{2021ApJ...908..170G} and \cite{2022AJ....164..195F}, the latter implementing a novel deep learning approach. We developed our own code which adapts the calculation of the Directional Light Radius (DLR; \citealt{2016AJ....152..154G}) to the Rubin Data Products and produces cutouts through the RSP to evaluate the results. The DLR is the elliptical radius of a galaxy in the direction of the SN in units of arcseconds. This metric takes also the extension and the orientation of the galaxies into account, and produces more reliable physical associations than using the plain angular separation. The DLR method, using only catalogue information, is fast and scalable to all kinds of extragalactic transients, as it makes no assumptions on its nature. Our code is publicly available and ready to work with Rubin Data Products on the RSP\footnote{\hyperlink{https://github.com/vpetrecca/Rubin-DP0-host-association}{github.com/vpetrecca/Rubin-DP0-host-association}}.

For each SN, we select host galaxy candidates in a region of $30''$ radius around the SN coordinates from the catalogue of static sources detected on stacked images (i.e., the LSST Object table). Such a radius is typically used in literature and is big enough to include the hosts of extragalactic transients. The full catalogue resulting from the query includes $277\ 852$ galaxies (hereafter, \textit{candidates}). Such a sample is also used in Sect. \ref{sec:photoz} to assess the quality of photometric redshifts, as it is big enough to include all typical DP0 galaxies. 

We assume the isophotes of the galaxy are self-similar ellipses and get the Stokes parameters Q and U from the adaptive second moments of source intensity $I_{xx}$, $I_{yy}$, $I_{xy}$:
\begin{equation}
    Q = I_{xx}-I_{yy}
\end{equation}
\begin{equation}
    U = I_{xy}
\end{equation}
The position angle (east of north) $\phi$ and the axis ratio $A/B$ of the galaxy are then
\begin{equation}
    \phi = \frac{1}{2}\arctan \left( \frac{U}{Q} \right)
\end{equation}
\begin{equation}
    A/B = \frac{1+k+2\sqrt{k}}{1-k}
\end{equation}
where $k$ is derived from the Stokes parameters as
\begin{equation}
    k \equiv Q^2 + U^2 = \left( \frac{A-B}{A+B} \right)^2
\end{equation}
We then measure the position angle $\alpha$ of the SN relative to the candidate host galaxy and we combine it with the angle $\phi$ to get the angle $\theta$ that the SN makes with the semi-major axis of the galaxy. 
Using the equation of a ellipse in polar coordinates with the origin at the center of the galaxy, we get the elliptical radius of a galaxy in the direction $\theta$, which is the definition of the DLR:
\begin{equation} \label{eq:DLR}
    r(\theta) = \frac{A}{\sqrt{(A/B\sin{\theta})^2+(\cos{\theta})^2}} \equiv DLR
\end{equation}
As an estimate of the semi-major axis $A$, we use a value of 2.5 times the Kron radius, that includes > 96\% of the total flux of a galaxy (\citealt{1980ApJS...43..305K}).
Finally, we define a dimensionless quantity by normalizing the angular separation $d_{sep}$ between the SN and the candidate host by DLR:
\begin{equation}
    d_{DLR} \equiv \frac{d_{sep}}{DLR} = \frac{d_{sep}}{A}\ C
\end{equation}
where $C$ is the denominator of Eq. \ref{eq:DLR}. In this way, we define a quantity which weights the SN radial distance by taking into account both the extension (the Kron radius, $A$) and the orientation (the $C$ parameter) of the galaxy. 

\begin{figure}
    \includegraphics[width=1.0\columnwidth]{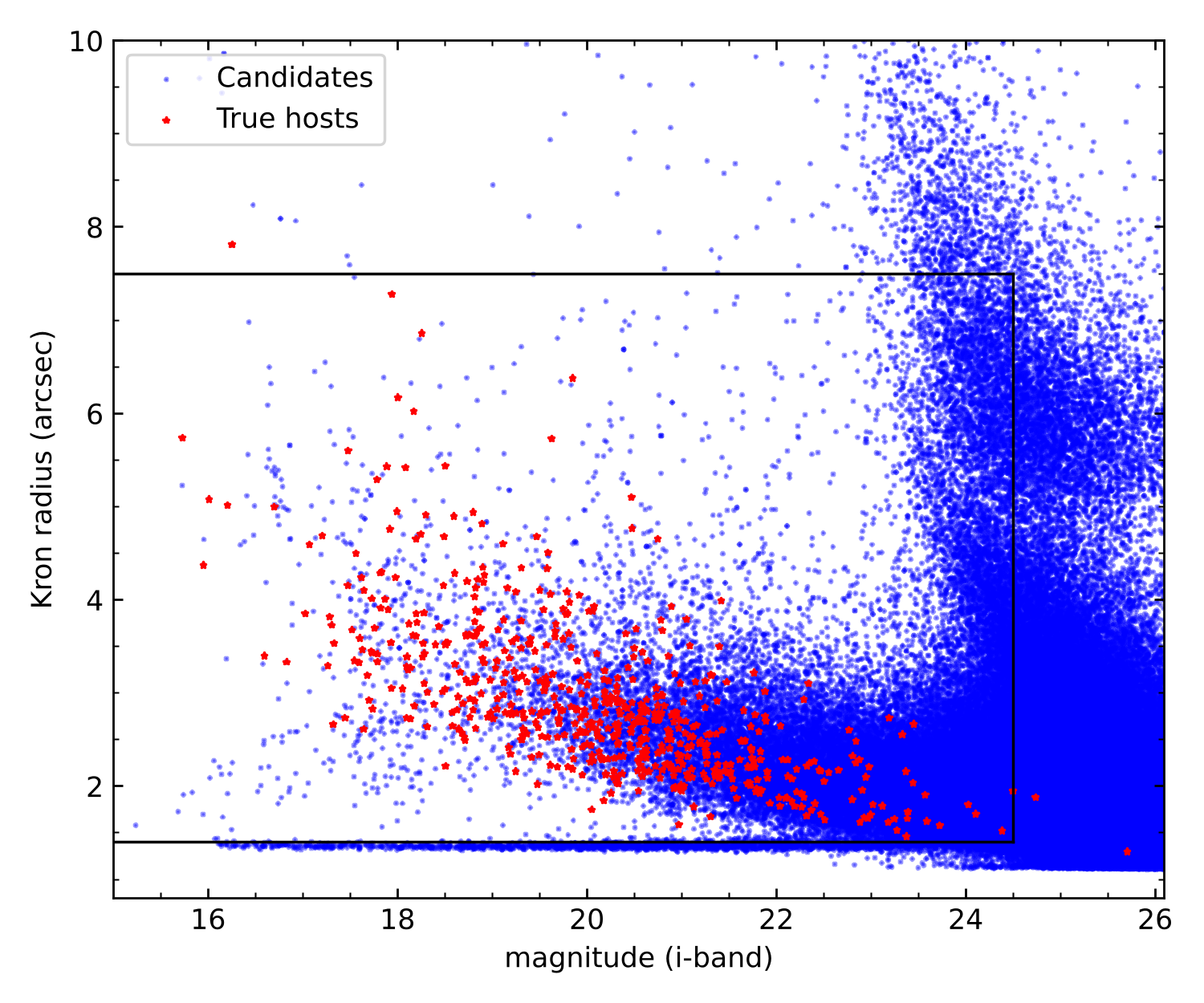}
    \caption{Kron radii and i-band magnitudes for the entire sample of host candidate (blue dots) and true (red stars) SN host galaxies. The black square determines the region actually used for the host association, avoiding faint galaxies and catastrophic estimates of the Kron radii.}
    \label{fig:kron_mag}
\end{figure}

For each SN, we rank all candidate galaxies by increasing $d_{DLR}$ and pick the one with the minimum $d_{DLR}$ as the best host. If minimum $d_{DLR} > 4$, we define the SN as host-less (SNe originally simulated as host-less in DC2 had been removed from the sample). As a further cut, we rejected candidate galaxies with \textit{i-band} magnitudes $mag_i \ge 24.5$ and with bad-quality fitting flags. These cuts reduce misassociations due to badly estimated galaxy parameters, mostly happening for low S/N sources. By looking at the distributions of candidates parameters, we also reject galaxies with Kron radii $\ge 7.5''$ or Kron radii $\le 1.4''$ to reduce misassociations due to catastrophic estimates of the galaxy extension (see Fig. \ref{fig:kron_mag}). All these cuts have an impact of only 1\% on the true host galaxy sample. We point out that the DLR method is strongly dependent on the quality of galaxy fitting. The Kron radii used in this work are computed using version 23 of the LSST Science Pipelines. Newer versions may improve the quality of galaxy parameters and require different cuts.

\begin{figure*}
    \begin{center}
    \includegraphics[width=0.33\linewidth]{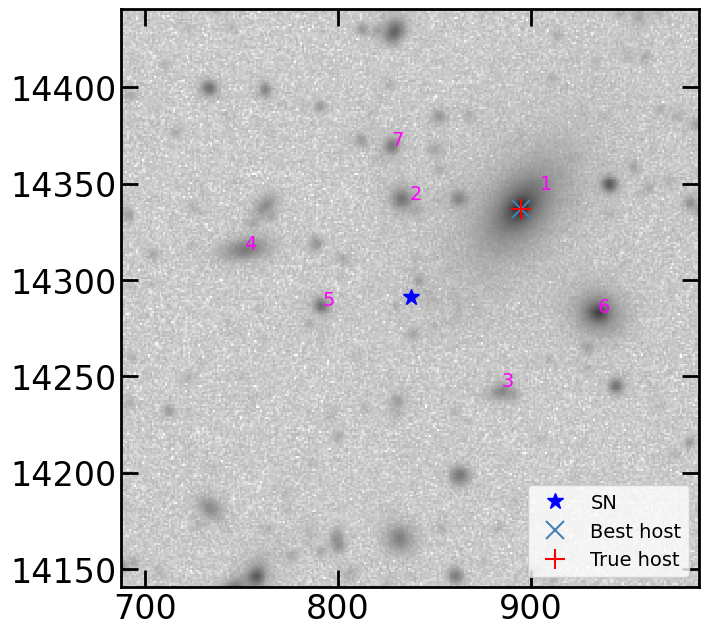}
    \includegraphics[width=0.33\linewidth]{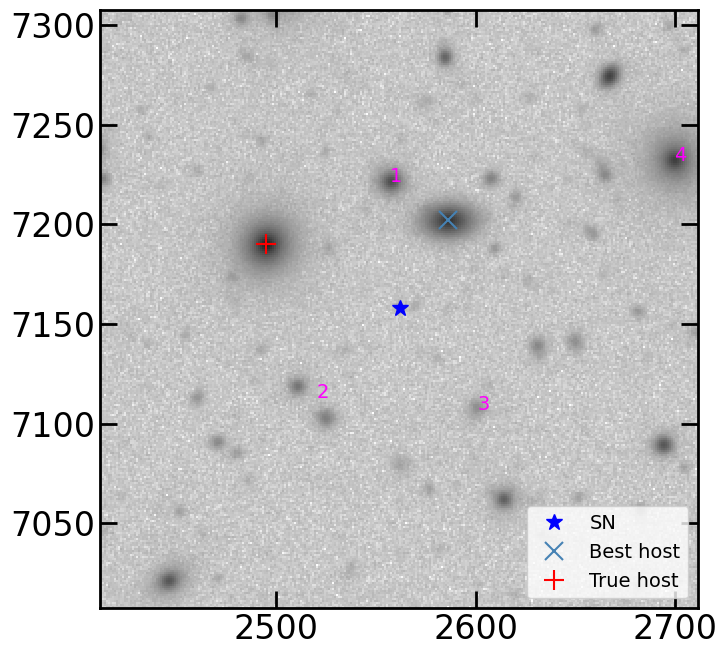}
    \includegraphics[width=0.33\linewidth]{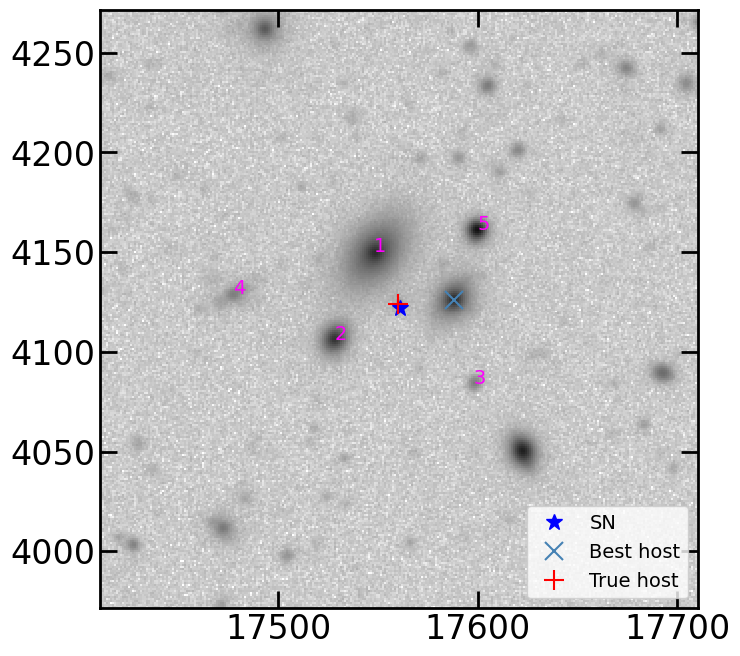}
    \end{center}
    \caption{Cutouts around SNe (blue star) extracted with the RSP. The magenta numbers identify possible host galaxies (ranked according to the lowest $d_{DLR}$), the blue x represents the best host candidate (the zeroth galaxy in the ranked list), while the red cross \textit{+} is the true host galaxy of the simulation. From left to right there are i) an example of correctly associated SN, and two possible misassociation cases: ii) very similar galaxies in terms of morphology and angular separation from the SN; iii) true host galaxy extremely faint and not among the possible list of candidates.} 
    \label{fig:hosts}
\end{figure*}

By comparing the results of our algorithm with the simulation input, we find that 89\% of SNe are correctly associated to the true host. Among the misassociations, 8 (i.e., $\sim 1\%$ of the total, 12\% of the wrong) are recovered as host-less, 35 (55\% of the wrong cases) are faint galaxies with $mag_i > 22$. The remaining 33\% is a result of projection issues (i.e., a bright galaxy along the line of sight of a SN actually injected into a faint host), or two similar galaxies in terms of morphology and angular separation from the SN, producing a similar $d_{DLR}$. Cutouts with examples of associations are reported in Fig. \ref{fig:hosts} for a correct case and two misassociations from the cases described above.

\begin{figure}
    \includegraphics[width=1.0\columnwidth]{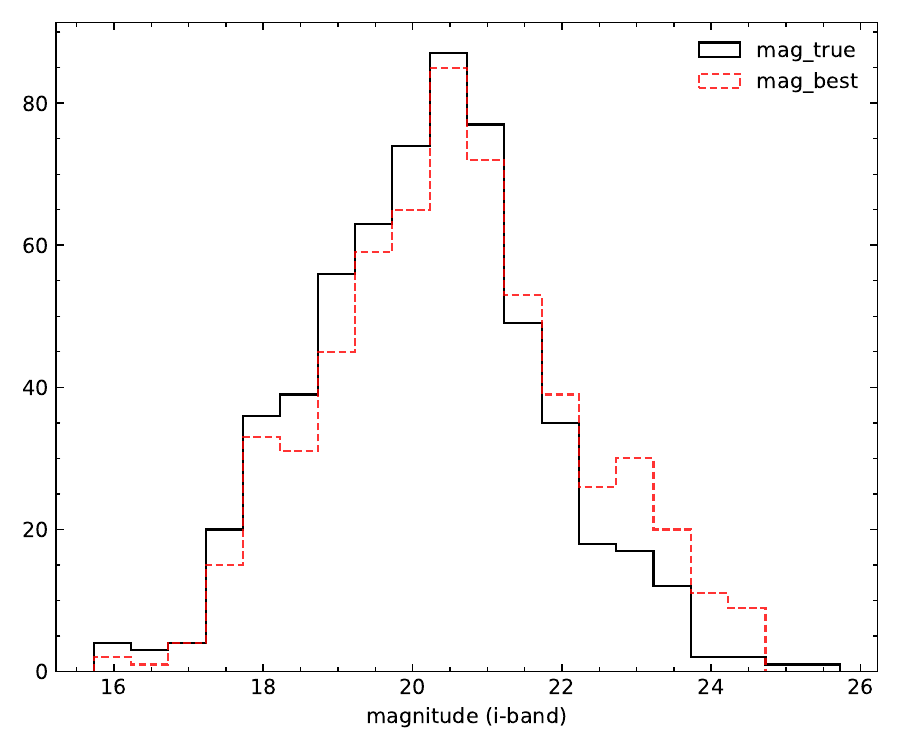}
    \caption{Distribution of i-band magnitude for the SN Ia host candidates. The black solid line is the magnitude of the true host galaxy (i.e., simulation input), while the red dotted line denotes the hosts associated with the DLR method.}
    \label{fig:mag_hist}
\end{figure}

Misassociations may also alter the distribution of the parent galaxy properties, affecting the final measurement of the SN rate. Figure \ref{fig:mag_hist} shows how the magnitude distribution of the SN hosts changes moving from the true to the associated host galaxies (hereafter denoted with \textit{"\_best"}). While there is an overall agreement for bright sources, it is worth noticing an excess of galaxies with $mag_i > 22.5$ for those associated with the DLR method. This will also have an effect on the accuracy of photometric redshifts, which tends to be lower with fainter sources as explained in Sect. \ref{sec:photoz}.

The association using only the minimum angular separation has a lower efficiency of 81\%. This highlights the importance of using more complex approaches as the DLR to associate transients to the host galaxies. Even better results could be obtained by considering correlations between the SN type and the galaxy properties, which are not included in this simulation.

\section{Photometric redshifts} \label{sec:photoz}

We cross-matched both the catalogues of true and associated host galaxies with a catalogue of photometric redshifts (zphot) for DC2 galaxies produced by \cite{2020MNRAS.499.1587S} with the BPZPipe code\footnote{\hyperlink{https://github.com/LSSTDESC/BPZpipe}{github.com/LSSTDESC/BPZpipe}}, which employs a training set complete up to $mag_i \le 25.0$. We used the weighted mean value of the posterior probability density function (PDF) for each galaxy as a point estimate of the photometric redshift.

\begin{figure}[h]
    \includegraphics[width=1.0\columnwidth]{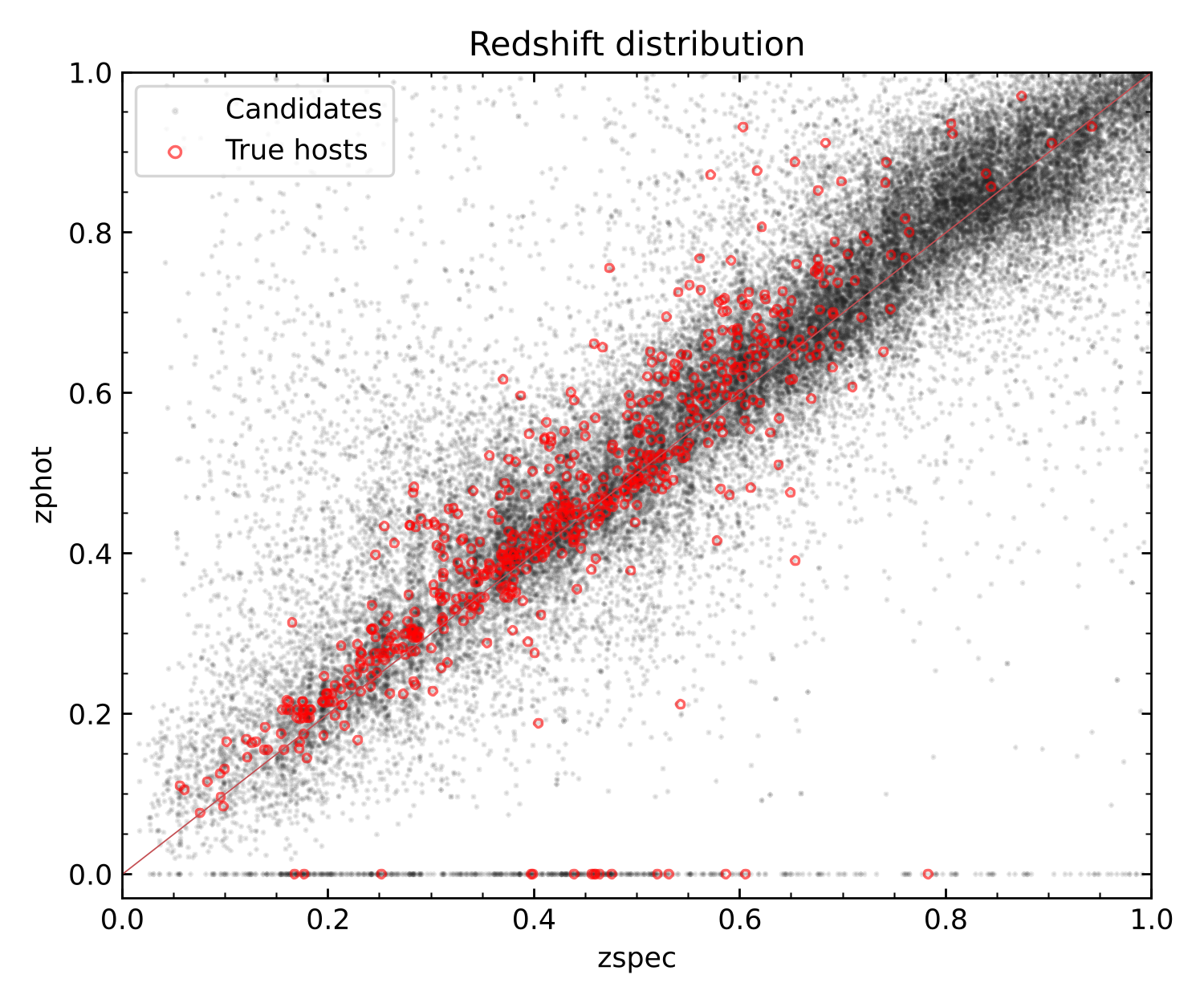}
    \caption{Distribution of the true simulated redshift (zspec) and photometric redshifts (zphot) for a sample of galaxies with $mag_i \le 25$. Black dots are for all host candidates, red open circles are the true SN Ia hosts.}
    \label{fig:z_dist_sample}
\end{figure}

To assess their quality, we used the sub-sample of all candidates host galaxies with $mag_i \le 25$ and $z_{spec} \le 1.0$, composed of $43161$ objects. By following the approach of \cite{2018AJ....155....1G}, we define a zphot error as $\Delta z = (z_{spec}-z_{phot})/(1+z_{phot})$, where $z_{spec}$ is the true simulated redshift. The robust standard deviation of the zphot error is evaluated in the interquartile range (IQR) containing 50\% of the catalogue galaxies. The value is then divided by 1.349 to convert it to the standard deviation of a Gaussian distribution ($\sigma_{IQR}$). This returns a value of $\sigma_{IQR}=0.05$ on the catalogue of all galaxies, and $\sigma_{IQR}=0.03$ on the true host galaxies of the SN Ia. The fraction of outliers (i.e., with $\Delta z > 3\sigma_{IQR}$) is $7\%$ for the catalogue of all galaxy candidates and $5\%$ for the SN Ia hosts. The comparison between the simulated and photometric redshift for both the sample of all candidate hosts and the true hosts is shown in Fig. \ref{fig:z_dist_sample}, where the better agreement for the true hosts is clearly evident as they are generally massive and brighter (see Fig. \ref{fig:kron_mag}). The horizontal line at the bottom of the plot identifies catastrophic outliers $z_{phot}\le 10^{-5}$, and 15 of them refer to true SN Ia hosts. Through all this work, we will adopt a fixed value of uncertainty of 0.05 on each photometric redshift estimate, as coming from the analysis of $\sigma_{IQR}$ on the galaxy catalogue. The value is also consistent with analysis of the median FWHM of the peak in the zphot PDFs.

\begin{figure}[h]
    \includegraphics[width=1.0\columnwidth]{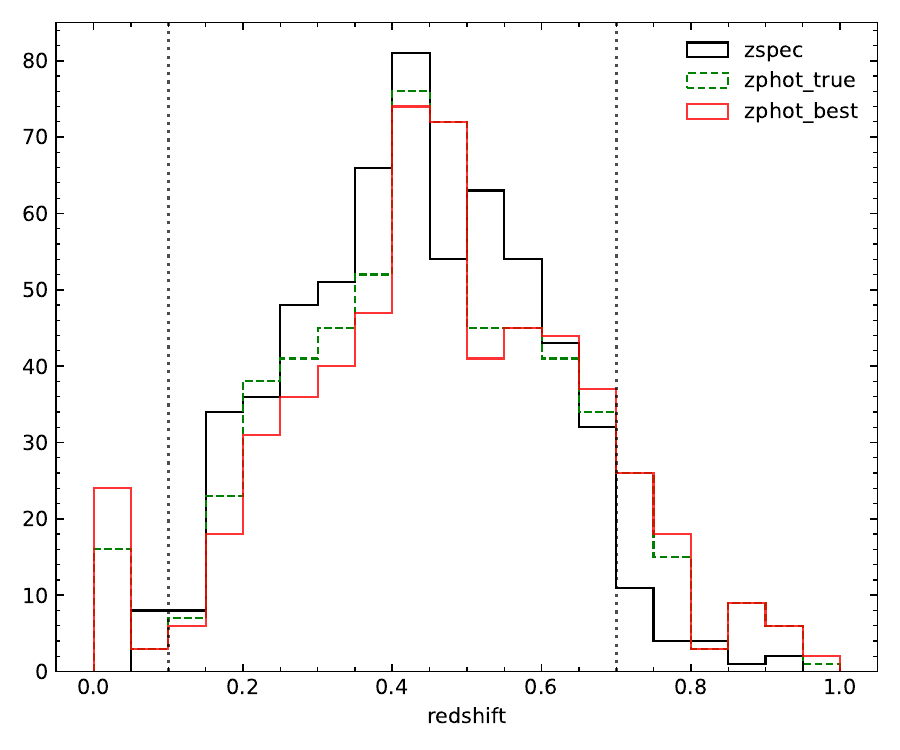}
    \caption{Distribution of redshift for the SN Ia host candidates. The black one uses the true simulated redshift (zspec), the green dashed line represents the photometric redshift of the true host galaxy (zphot\_true), while the red continuous distribution refers to the photometric redshift of the associated host galaxy (zphot\_best). The two vertical lines define the range where we evaluate the volumetric rate and the plot is cut to $z = 1.0$ for visual clarity.}
    \label{fig:z_hist}
\end{figure}

Figure \ref{fig:z_hist} shows the combined impact of photometric redshift and host galaxy association on the retrieved SN sample redshift distribution. As expected, the redshift distribution of the SN hosts appears wider when using the photometric redshift of the associate host (z\_phot\_best). This is especially true for faint galaxies whose uncertainty on photometric redshift is higher. We notice a peak at $z_{phot} \sim 0$ due to catastrophic outliers or SNe wrongly associated to nearby galaxies, and a peak at $z_{phot} \sim 0.45$ mainly due to the known degeneracy between the Lyman break of galaxies at higher redshift and the Balmer break of galaxies at lower redshift (\citealt{2001A&A...380..425M}; \citealt{2019NatAs...3..212S}). These peaks, together with an excess of sources at higher redshifts, result in a decrease of SNe in all the other redshift bins. 

In Sect. \ref{sec:rate} we will quantitatively discuss the effect of the changing distribution on the volumetric rate by considering redshift bins of width 0.05 (which is the nominal error on photometric redshifts). The analysis is restricted to the redshift range $0.1 \le z \le 0.7$ to avoid a catastrophic drop of simulated SNe at the edges of the distribution. As shown in Fig. \ref{fig:z_hist}, we expect a loss of SNe in all the redshift bins with an excess at $z_{phot} \sim 0.45$. We also point out that, in principle, all the effects of bin migration arising from this work could be taken into account with a Monte Carlo analysis leading to correction coefficients (see e.g., \citealt{Lasker}). However, estimating the proper corrections taking into account all sources of uncertainty (including those related to the detection) would require producing more DP0-like simulations. The present work focuses only on what is provided in DP0.2, as a single realisation of the Universe processed with the LSST science pipelines. The production of more complete simulations is under investigation for future works. 

\section{Classification}\label{Sec:class}

For large surveys like LSST, the photometric classification of transients is an essential task. Although the brokers will instantly process the nightly alert streams\footnote{\hyperlink{https://www.lsst.org/scientists/alert-brokers}{lsst.org/scientists/alert-brokers}}, this will mainly serve as a pre-classification to enable follow up of specific sources of interest with other facilities. More thorough analysis may require a posteriori classification using all the available information.

In this section, we present the results of the classification with three different approaches, based on light curves of the SN Ia sample. The first two methods involve template fitting and are part of the publicly available \textbf{S}uper\textbf{N}ova \textbf{ANA}lysis software package (\texttt{SNANA}\footnote{\hyperlink{https://github.com/RickKessler/SNANA}{https://github.com/RickKessler/SNANA}}; \citealt{2009PASP..121.1028K}), while the last one uses a recurrent neural network (RNN) trained on light curves. For all the methods, we obtain a classification: i) using only light curves, ii) providing the photometric redshift of the associated host galaxy as a prior (if available\footnote{SN Ia identified as host-less or having catastrophic photometric redshifts (i.e., $z_{phot}\le 10^{-5}$) are classified with a flat prior on redshift.}), iii) providing the true simulated redshift as a prior, which is the case when we have the spectroscopic redshift of the true host galaxy from other surveys. 

The final results of the classification, using all the methods, are summarized in Table \ref{tab:summary_class}. Although our classification models have been trained on multiple classes, the presence of SN Ia only in the DC2 simulation does not allow us to measure the actual effect of contamination from other transients (especially from SN Ib/c). The mis-classification effect will then only be a reduction of the total number of SN Ia from the original sample. We refer to upcoming works analyzing the \texttt{ELAsTiCC} light curves for a deep overview of photometric classification. 

\subsection{PSNID} \label{sec:class:psnid}

The \textbf{P}hotometric \textbf{S}uper\textbf{N}ova  \textbf{ID}entification (\texttt{PSNID}; \citealt{2011ApJ...738..162S}) is a template fitting algorithm which calculates the reduced $\chi^2$ against a grid of SN Ia light curves models and core-collapse SN templates in order to identify the best-matching SN type. It has been firstly used for prioritizing spectroscopic follow-up observations for the SDSS-II SN Survey (\citealt{2008AJ....135..348S}), and it has been also tested with \texttt{SNANA} simulations from the Supernova Photometric Classification Challenge (\citealt{2010arXiv1001.5210K}). 

The version we used is integrated in \texttt{SNANA} and computes also Bayesian probabilities for different SN types. More into detail, the Bayesian evidence $E$ for each SN type is calculated by marginalizing the product of the likelihood function and prior probabilities over the model parameter space:
\begin{equation}
    E_{type} =  \sum_{template} \int_{pars.\ range} P(z)e^{-\chi^2/2} dz dA_V dT_{max} d\mu 
\end{equation}
Fitting parameters are redshift $z$, with probability distribution $P(z)$, extinction $A_V$, time of maximum $T_{max}$ and distance modulus $\mu$. In this work, priors in $A_V$, $T_{max}$ and $\mu$ are flat, while for redshift we tested both a flat prior with $P(z)=1$, and a Gaussian prior by providing the mean and sigma as described in Sect. \ref{sec:photoz} when using the photometric redshifts, and a sigma of 0.0001 when using the spectroscopic prior. For the SN Ia light curves we used the SALT2-Extended model (\citealt{2018PASP..130k4504P}), the same that had been used in the simulation, while templates for SNe core-collapse come from \cite{2019MNRAS.489.5802V}. SN template light curves are K-corrected using tables produced for LSST by DESC.

\begin{figure*}
    \begin{center}
    \includegraphics[width=0.49\linewidth]{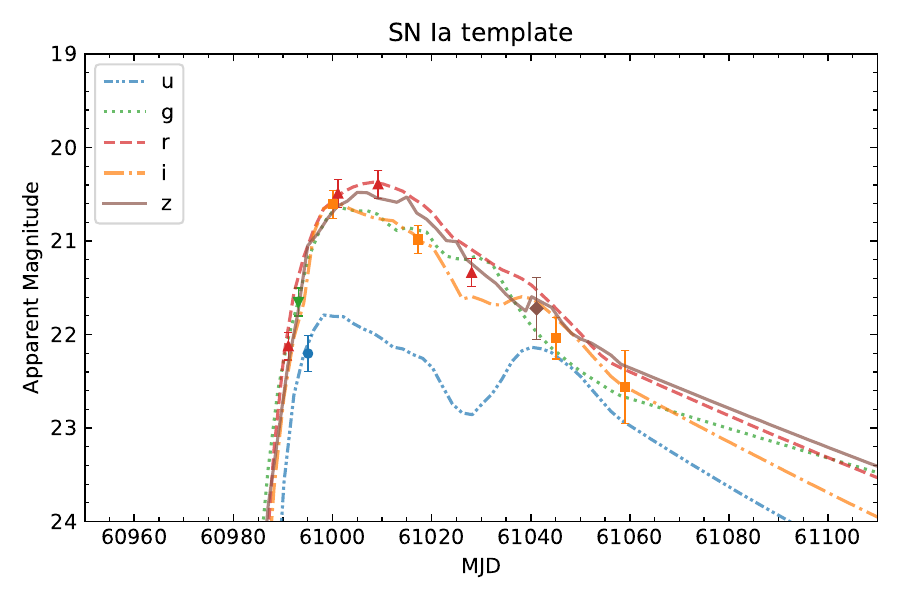}
    \includegraphics[width=0.49\linewidth]{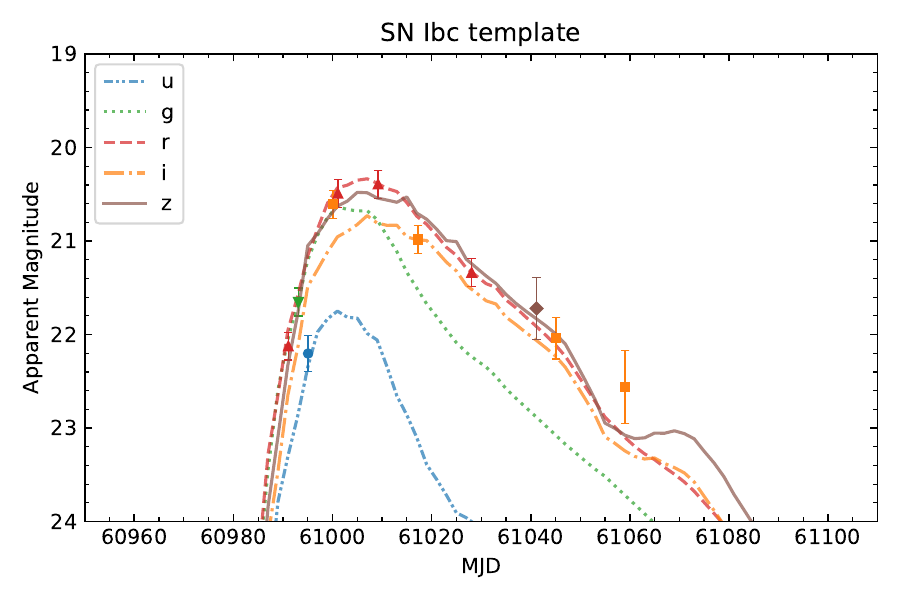}
    \end{center}
    \caption{Example of template fitting with \texttt{PSNID} for a SN Ia and a SN Ib/c model on the same light curve. Different colors and lines refer to different photometric bands. Left panel shows the fitting with the best Ia template, while the right panel shows the best-fitting SN Ib/c template. Although both fits seem reasonable, not allowing a straightforward classification, the type Ia has a Bayesian probability $P_{Ia}=1.0$ with the redshift prior.}
    \label{fig:psnid_fits}
\end{figure*}

The Bayesian probability for each SN type is then defined as
\begin{equation}
    P_{type} = \frac{E_{type}}{E_{Ia}+E_{Ib/c}+E_{II}}
\end{equation}
where
\begin{equation}
    P_{Ia} + P_{Ib/c} + P_{II} = 1.
\end{equation}
With both Bayesian probability and $\chi^2$ available, there are many ways of selecting a sample of SN Ia. \texttt{PSNID} comes with a default criterion which takes both of them into account and defines the SN as of "unknown" type if the difference between the $\chi^2$ of two SN types is smaller than a certain threshold. However, as discussed in \cite{2007ApJ...659..530K}, the minimum $\chi^2$ alone is not always a good indicator of the best SN type. This become even worse when small photometric errors lead to unreasonably high values of $\chi^2$. While the issue is often mitigated by artificially increasing the errors on light curves, we preferred to rely only on the Bayesian probability and selected all SNe with $P_{Ia} > 0.5$. With this conservative approach, we maximize the completeness of the sample, which is very important for the evaluation of the SN rate. Figure \ref{fig:psnid_fits} shows an example of template fitting resulting in a SN Ia according to the Bayesian probability, but of unknown type when using the $\chi^2$.

The fractions of SNe correctly classified as SN Ia are very high when the redshift prior is used (95\% with the associated zphot and 99\% with the true simulated redshift). However, without a prior on redshift, only 44\% of SNe are correctly classified, and the resulting redshift distribution is peaked towards lower values. This occurs because of a degeneracy between extinction and redshift in the light curves reddening and it shows the importance of using priors, when available.

\subsection{SALT2 fitting} \label{sec:class:salt2}

Another approach often used in SN Ia cosmology consists of fitting a selected model to the light curves and taking only those candidates providing good fits (e.g., \citealt{2022ApJ...934...96S}; \citealt{2023ApJ...944..212M}). The resulting sample will have high purity, which is essential for cosmological analysis, but the procedure only allows to select sources conforming to the predefined model. 

As our sample contains only SN Ia, we tested this methods using the software \textit{snlc\_fit.exe} distributed by \texttt{SNANA} and assuming again the SALT2-Extended model. The best fit is determined through a minimization procedure based on CERNLIB's \texttt{MINUIT} program\footnote{\hyperlink{https://root.cern.ch/doc/master/classTMinuit.html}{root.cern.ch/doc/master/classTMinuit.html}} and two iterations on each light curve. The fitting requirements are very similar to those adopted in \cite{2022ApJ...934...96S} on the same dataset: 
\begin{itemize}
    \item stretch parameter $|x_1| < 3$;
    \item color parameter $|c| < 0.3$;
    \item fit probability ($P_{fit}$) computed from $\chi^2$ and number of degrees of freedom satisfying $P_{fit}>0.05$.
\end{itemize}
 Similarly to what we did with \texttt{PSNID}, we tested the SALT2 fit both with a Gaussian prior on redshift and with no redshift prior.

The fitting procedure results in a selection efficiency of 75\% when the photometric redshift of the associated galaxy is used as a prior, and 69\% with no redshift information. The success fraction for the no-zphot prior scenario when using SALT2 fitting is higher than that of \texttt{PSNID}. The main reason is that \texttt{PSNID} is doing a more complicated classification between different classes, while here we are just fitting a single model to all our sources.

\subsection{SuperNNova} \label{sec:class:supernnova}
Recent advantages in deep neural networks make them extremely promising to photometrically classify variable sources for large surveys like LSST. Indeed, they can be trained on both simulated and archival data, enabling a fast multi-class analysis not limited either by costly feature extraction or by template matching biases. 
\texttt{SuperNNova}\footnote{\hyperlink{https://github.com/supernnova/SuperNNova}{github.com/supernnova/SuperNNova}} (SNN; \citealt{2020MNRAS.491.4277M}) is an open-source framework requiring only photometric time-series as input, with additional information (e.g., host galaxy redshift) that can be provided to improve performances. SNN includes different classification methods, as long short-term memory (LSTM; \citealt{lstm}) recurrent neural networks (RNNs) and two Bayesian neural networks (BNN). 

As with the previous approaches, we tested SNN on our SN Ia sample both with and without redshift information using the default configuration RNN. We trained our models using a sample of synthetic light curves of all SN type with LSST like photometric performances from \texttt{ELAsTiCC}. The simulation also includes spectroscopic and photometric host galaxy redshifts for $\sim 5$ million objects, thus allowing to build training sample with different redshift priors. We trained 3 different models for a binary SN Ia vs non-Ia classification: i) light curves only, ii) with host galaxy photometric redshift, iii) with spectroscopic redshift. 
The SNN output is a probability of being SN Ia, $P_{Ia}$, for each SN event. We considered as correctly classified all SN with $P_{Ia} > 0.5$. This resulted in 96\% or 97\% accuracy when the photometric or spectroscopic redshift information is used (compatible with \texttt{PSNID}). However, without redshift prior, only 80\% of SNe is correctly classified as SN Ia. 

The lower efficiencies for all the methods in absence of redshift information show how fundamental it is to have reliable estimates of photometric redshifts and a good host association procedure. This is especially true if the number of detections and their distribution around the light curve peak is not sampled enough to enable a proper classification (see e.g., \citealt{2022ApJS..258...23A}).

\begin{table}[httb]
\centering
\begin{tabular}{c c c c} 
 {Method} & {no z prior} & {zphot\_best prior} & {zspec prior} \\
 \hline
 {PSNID} & {44\%} & {95\%} & {99\%} \\
 {Salt2 fit} & {69\%} & {75\%} & {80\%} \\
 {SuperNNova} & {80\%} & {96\%} & {97\%} \\
 \hline\
\end{tabular}
\caption{Fraction of correctly classified SNe from the SN Ia sample using different methods.}
\label{tab:summary_class}
\end{table}


\section{SN Ia Rate} \label{sec:rate}

The measurement of the SN rate requires accurate information on the survey strategy and design, both for what concerns the detection efficiency and the observing cadence. The former is usually determined with simulations and injection of point-like sources onto images, exploring a wide range of magnitudes and positions on sky (\citealt{2015A&A...584A..62C}). The impact of the latter is traditionally measured with the \textit{control time} (\citealt{1942ApJ....96...28Z}), defined as the interval of time during which a SN occurring at a given redshift is expected to remain above the detection limit of the image.

However, the DC2 observing strategy is not the definitive version that will be adopted for LSST. Moreover, the main aim of this work is the analysis of the impact on the rate of sources of uncertainties that are not directly related to the survey strategy. For this reasons, we adopted a simplistic approach consisting in evaluating a single \textit{recovery efficiency} term, which takes into account both control time and detection efficiency, from the simulated rate. We refer to other works for a thorough analysis of detection efficiency and magnitude limits on DC2 simulation (\citealt{2022ApJ...934...96S}) and the impact of different observing cadence on the classification of SNe (\citealt{2022ApJS..258...23A}).

\begin{figure}
    \includegraphics[width=1.0\columnwidth]{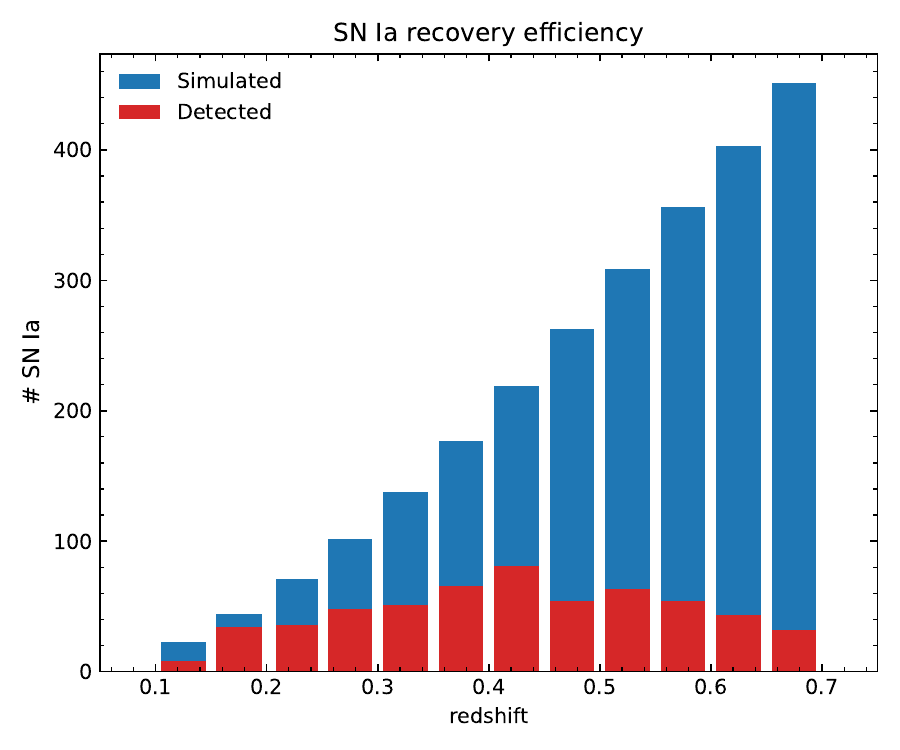}
    \caption{Numbers of SN Ia simulated (blue) and detected (red, the SN Ia sample) in the redshift bins used for the measurement of the rate.}
    \label{fig:rate_eff2}
\end{figure}

We considered the SN Ia sample in the range $0.1 \le z_{spec} \le 0.7$ (as explained in Sect. \ref{sec:photoz}), and divided it in redshift bins of width 0.05, which is the typical error of a photometric redshift measurement. The selected range of redshift contains 570 of the 600 SN Ia in the SN Ia sample. We define the \textit{recovery efficiency} as the ratio of the number of detected sources in the SN Ia sample to the total number of simulated SN Ia in each redshift bin over an observing window of $T=3.5$ years, which is the effective observing time removing gaps between seasons. The distributions with the number of simulated and detected SN Ia in the redshift bins is showed in Fig. \ref{fig:rate_eff2}, while the ratios leading to the \textit{recovery efficiency} $\epsilon$ are reported in Table \ref{tab:efficiency}.

\begin{table}
\centering
\begin{tabular}{c c c c} 
 {Simulated} & {Detected} & {Bin} & {$\epsilon$} \\
 \hline
 {23}  & {8}  & {$0.10 \le z < 0.15$} & {34.8\%} \\
 {44} & {34}  & {$0.15 \le z < 0.20$} & {77.3\%} \\
 {71} & {36} & {$0.20 \le z < 0.25$} & {50.1\%} \\
 {102} & {48} & {$0.25 \le z < 0.30$} & {47.1\%} \\
 {138} & {51} & {$0.30 \le z < 0.35$} & {37.0\%} \\
 {177} & {66}  & {$0.35 \le z < 0.40$} & {37.3\%}  \\
 {219}  & {81}  & {$0.40 \le z < 0.45$} & {37.1\%} \\
 {263} & {54}  & {$0.45 \le z < 0.50$} & {20.6\%} \\
 {309} & {63} & {$0.50 \le z < 0.55$} & {20.4\%} \\
 {356} & {54} & {$0.55 \le z < 0.60$} & {15.2\%} \\
 {403} & {43} & {$0.60 \le z < 0.65$} & {10.1\%} \\
 {451} & {32}  & {$0.65 \le z \le 0.70$} & {7.1\%}  \\
 \hline\
\end{tabular}
\caption{SN Ia recovery efficiency $\epsilon$ in different redshift bins.}
\label{tab:efficiency}
\end{table}

Depending on the adopted estimate of redshift, the distribution of SNe is different, as showed in Fig. \ref{fig:z_hist}. The total number of recovered SN Ia when looking at photo-z is lower than those in the simulated sample because of sources being assigned a redshift outside the selected range. Moreover, additional SN Ia are lost because of mis-classifications and the absence of non-Ia contaminants. In this section, we evaluate (separately) the uncertainty contribution of photometric redshifts, host galaxy association and photometric classification on the measurement of the SN Ia rate. As a reference classification tool, we use \texttt{PSNID} adopting the zphot\_best prior of the associated galaxy. The reasons for this choice are the flexibility of the algorithm in dealing with different SN types, the possibility to customize the output by defining an "unknown" class, and the fast execution (intrinsically parallelizable) without requiring to build new training sets. The classification accuracy of the other multi-class algorithm (\texttt{SuperNNova}) is similar, thus the choice of the classification method does not affect the final result. 

\begin{figure}
    \includegraphics[width=1.0\columnwidth]{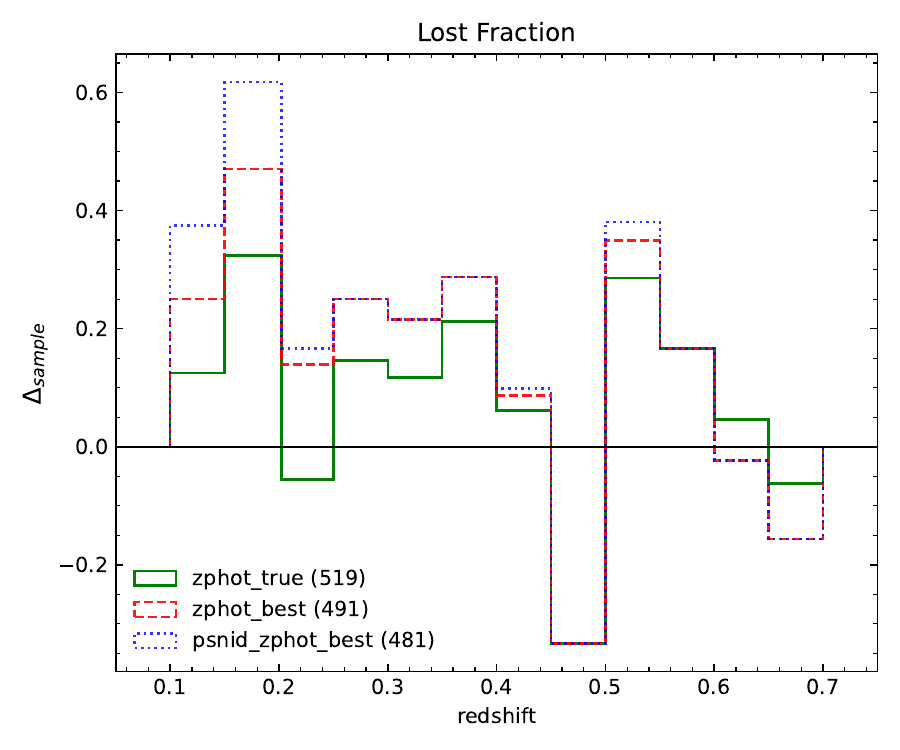}
    \caption{Lost fraction of SNe in each redshift bin for the different samples, highlighting the impact of photometric redshift \textit{(zphot\_true)}, host galaxy association \textit{(zphot\_best)}, and classification \textit{(psnid\_zphot\_best)}. The numbers in parenthesis refer to the total number of SNe in each subset for the selected redshift range. Negative values of $\Delta_{sample}$ denote an increase of sources in that redshift bin (see Eq. \ref{eq:delta}).}
    \label{fig:rate_missing}
\end{figure}

The samples used in our analysis are the following:
\begin{itemize}
    \item \textbf{\textit{zphot\_true}:} SN Ia sample adopting the photometric redshift of the true host galaxy instead of the true simulated redshift (this allows to test the effect of photometric redshift);
    \item \textbf{\textit{zphot\_best}:} SN Ia sample adopting the photometric redshift of the associated host galaxy (effect of photometric redshift + host galaxy association);
    \item \textbf{\textit{psnid\_zphot\_best}:} sample of SNe correctly classified as SN Ia adopting the photometric redshift of the associated host galaxy as a prior (effect of photometric redshift + host galaxy association + classification uncertainties).
\end{itemize}
A first measurement of the contribution of the different effects comes from the lost fraction of SNe for the various samples $\Delta_{sample}$, providing an indication of the number of SN Ia lost because of wrong photometric redshift, host association and classification. We define it in each redshift bin as: 
\begin{equation}
    \Delta_{sample} = (N_{detected}-N_{recovered})/N_{detected},
    \label{eq:delta}
\end{equation}
where $N_{detected}$ refers to the original SN Ia sample (not affected by the uncertainties analyzed in this work) and $N_{recovered}$ is the number of SN Ia in each one of the samples defined above. The resulting values as a function of redshift are shown in Fig. \ref{fig:rate_missing} with different colors for the different subsets. As already noticed by looking at the redshift distribution in Fig. \ref{fig:z_hist}, there is an overall depletion of SNe with the exception of the bins at $z \sim 0.45$ and $z > 0.6$. The \textit{sample\_psnid\_zphot\_best}, which include all the effects together, has an average lost fraction of $17\%$. However, photometric redshift alone, measured with the \textit{sample\_zphot\_true}, has an impact of $10\%$ and results to be the major source of uncertainty (also affecting classification when used as prior). It is also worth noticing how the different effects depend on redshift: at $z > 0.4$ most of the mismatch comes from the error on the photometric redshift, while lower redshifts are more affected by wrong host galaxy associations and wrong photometric classification. 

We now want to test what is the effect of the different biases on the derived SN rates. To this aim, we compute the SN Ia volumetric rate for each sample as: 
\begin{equation}
    r_{SN Ia}(z) = \frac{(1+z)}{V(z)}\frac{N_{SN}(z)}{T\epsilon(z)},
\end{equation}
where $N_{SN}(z)$ is the number of SNe in each redshift bin for the selected sample, $T$ is the observing window, $(1+z)$ corrects for time dilation, $\epsilon(z)$ is the SN Ia \textit{recovery efficiency} defined in Table \ref{tab:efficiency}, and $V(z)$ is the comoving volume for the given redshift bin:
\begin{equation}
    V(z) = \frac{4\pi}{3}\frac{\Theta}{41253}\bigg[\frac{c}{H_0}\int^{z2}_{z1}\frac{dz'}{\sqrt{\Omega_M(1+z')^3+\Omega_{\Lambda}}}\bigg]^3\ Mpc^3.
\end{equation}
In the previous equation, $\Theta$ is the search area of $15\ deg^2$, $z$ is the mid-point of the redshift interval $[z1,z2]$, and we assumed a flat $\Lambda$CDM universe with $H_0=70$ and $\Omega_M = 0.3$.

\begin{figure}
    \includegraphics[width=1.0\columnwidth]{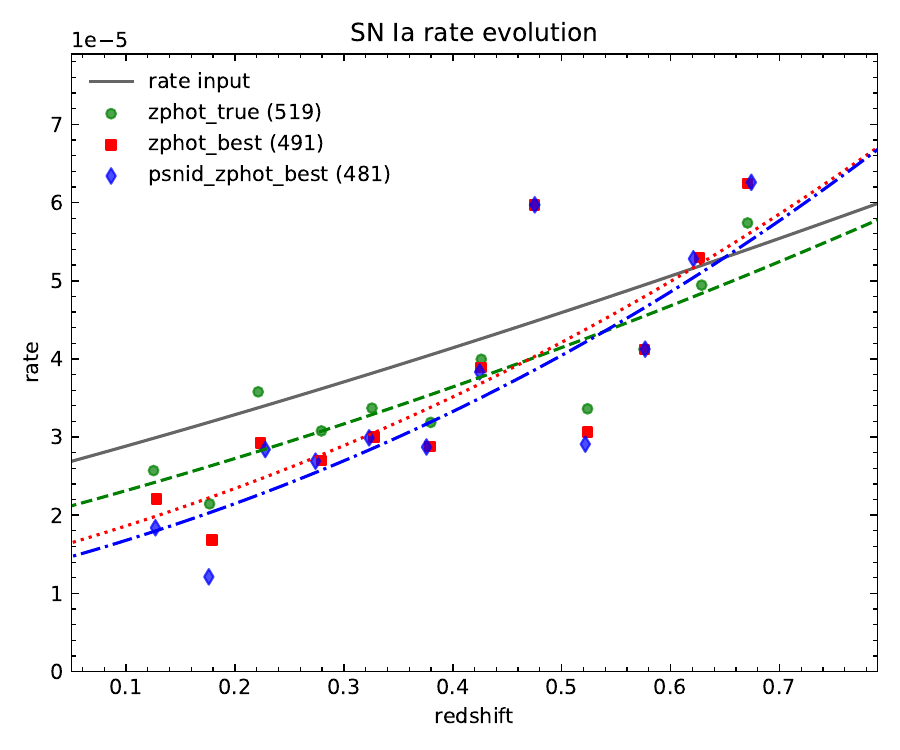}
    \caption{SN Ia rate for the different samples, as in Fig. \ref{fig:rate_missing}. Continuous lines are power-law fits of the rate points. Error bars due to statistical uncertainty are omitted for visual clarity.} 
    \label{fig:rate_fit}
\end{figure}

The resulting SN Ia rates for the different samples are shown in Fig. \ref{fig:rate_fit}, along with power-law fits with the same functional form of the simulated rate from \cite{2008ApJ...682..262D}:
\begin{equation}
    r_v(z) = \alpha\times10^{-5}(1+z)^{\beta}Mpc^{-3}yr^{-1}.
\end{equation}
The fit results are summarized in Table \ref{tab:rate}. The differences between the fit parameters and the input rate ($\alpha=2.5$ and $\beta=1.5$) show that the uncertainties not only reduce the number of SN Ia in each redshift bin, but also change the evolutionary trend of the recovered rate. The combination of the two effects hampers the use of the volumetric rate to discriminate between SN Ia progenitor models, unless the impact of these sources of uncertainty can be reduced. 

\begin{table}[b]
\centering
\begin{tabular}{c c c c c} 
 {Sample} & {$\alpha$} & {$\beta$} & {$\alpha_{err}$} & {$\beta_{err}$} \\
 \hline
 {zphot\_true} & {1.9}  & {1.9} & {0.3} & {0.4}\\
 {zphot\_best} & {1.4} & {2.6} & {0.3} & {0.4}\\
 {psnid\_zphot\_best} & {1.3} & {2.8} & {0.3} & {0.5}\\
 \hline\
\end{tabular}
\caption{Fit coefficients for all the samples. The input rate has $\alpha=2.5$ and $\beta=1.5$.}
\label{tab:rate}
\end{table}

\begin{figure}
    \includegraphics[width=1.0\columnwidth]{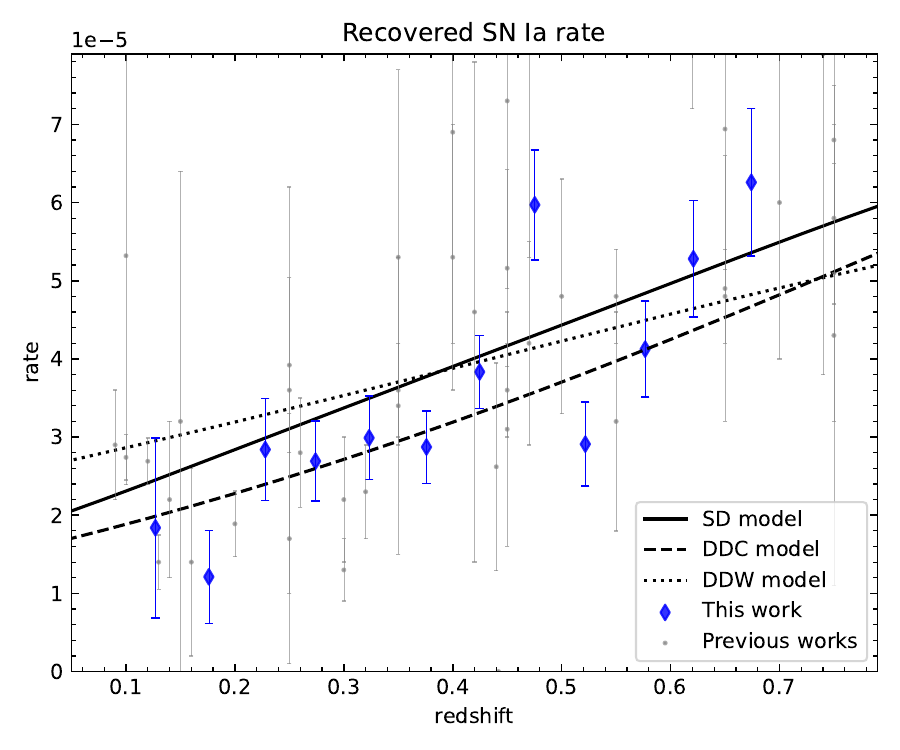}
    \caption{The recovered SN Ia rate for the \textit{sample\_psnid\_zphot\_best} (the real observing case) with rate predictions of progenitor models from \cite{2005A&A...441.1055G}. Error bars on the blue points are due to statistical uncertainties scaled to 10 years over the simulated $\sim 15\ deg^2$ DC2 universe (even though we expect higher statistics with the real survey). The grey points are the literature measurements shown in Fig. \ref{fig:rate_enrico}.}
    \label{fig:rate_models}
\end{figure}

To quantitatively address the problem, we focused on the \textit{sample\_psnid\_zphot\_best}, which represents a real case scenario combining all the uncertainties (i.e., a study of SN Ia rate with photometric data only). Figure \ref{fig:rate_models} shows the recovered rate along with predictions from the progenitor models described in Sect. \ref{sec:intro} (where the $k_{Ia}$ factor in Eq. \ref{eq:rate_models} has been fixed to $0.8\times 10^{-3} M_{\odot}^{-1}$ for all scenarios). Error bars on the blue point are due to statistical uncertainties scaled to 10 years over the simulated $\sim 15\ deg^2$ (even though we expect higher statistics with the real survey, with statistical errors reduced by more than one order of magnitude). The grey points represent rate measurement from the literature, as shown in Fig. \ref{fig:rate_enrico}. It is clearly evident how LSST will reduce the scatter between the rate estimates obtained comparing multiple surveys up to $z \sim 1.0$. However, biases introduced by other sources of uncertainty, such as those analyzed in this work, should still be reduced to attain optimal results. 
Despite Fig. \ref{fig:rate_enrico} showed that in the intermediate redshift range probed by LSST different progenitor models have similar outcomes, the combination of LSST and higher redshift surveys (e.g., the Nancy Grace Roman Space Telescope; \citealt{roman_sin}; \citealp{roman}) could indeed allow to distinguish between different scenarios. Moreover, \cite{2019A&A...625A.113G} show that the SN Ia rate as a function of host galaxy colors or specific SFR is even more efficient in separating the predictions of different models, once all the uncertainties have been reduced. 

\section{Summary and conclusions} \label{sec:concl}

The LSST is expected to increase the number of detected SN Ia by a factor of 100 compared to samples from previous surveys (\citealt{2019ApJ...881...19J}). This will dramatically reduce statistical uncertainties on the SN Ia rate measurement, possibly allowing us to put constraints on the SN Ia progenitors by comparing the observed rate with predictions from theoretical models. However the actual impact of all the possible sources of uncertainty on the measurement of the rate deserves further analysis. 
While many observational biases in the selection of a good sample of SN Ia for cosmology have already been inspected in other works, here we studied the uncertainties due to estimate of photometric redshift, host galaxy association and classification on the measurement of the SN Ia rate using simulated LSST images. 

Data come from \cite{2022ApJ...934...96S}, who executed DIA on a subset of $\sim 15\ deg^2$ of the DC2 simulation. There are a total of 5884 simulated SNe with $z \le 1.0$, 2186 of them detected on difference images. We selected only sources with more than 5 distinct detections in order to have a sufficiently sampled light curve for the transient classification. The analyzed SN Ia sample consists than of only 600 sources, an order of magnitude lower than the number of simulated SN Ia. The large loss of sources is mainly due to  the sub-optimal observing cadence of the simulated WFD region and testifies the need of heavily cadenced observations on smaller region of the sky for significant statistical studies. Indeed, the definition of the best cadence is still an open issue and there are many works using simulations to provide quantitative metrics (e.g., \citealt{2022ApJS..259...58L}).

We associate each SN Ia to the host galaxy using the DLR. Our algorithm has an association accuracy of 89\% using only morphological information extracted from a single band deep coadd image. Among the misassociation, 12\% is recovered as host-less, 55\% is associated to a faint galaxy with $mag_i > 22$, and the remaining 33\% is a combination of projection issues or two similar and close galaxies producing a similar $d_{DLR}$. The host association could be further improved with a better estimate of the photometric parameters for faint galaxies and by considering also the correlation between the SN type and the host galaxy properties (not included in the simulation). 

We recovered estimates of the SN photometric redshifts from both the true and the associated host galaxy. The quality of the photometric redshift has been studied with a sample of galaxies with $mag_i \le 25$ and $z_{spec} \le 1.0$. The analysis returned a robust standard deviation of $\sigma_{IQR}=0.05$ and the fraction of outlier is 7\%. The combined impact of photometric redshift and host galaxy association results in a broadening of the SN Ia distribution in redshift. We found an excess of associated host galaxies with $mag_i > 22.5$, along with a peak at $z_{phot} \sim 0$ (due to catastrophic outliers or SNe wrongly associated to nearby galaxies), and a peak at $z_{phot} \sim 0.45$ (mainly due to the known degeneracy between the Lyman break of galaxies at higher redshift and the Balmer break of galaxies at lower redshift). 

Light curves have been classified with different methods, involving both template fitting techniques (\texttt{PSNID} and SALT2 model fit by \texttt{SNANA}) and recurrent neural networks (\texttt{SuperNNova}). All the algorithms have been executed with and without prior on redshift, proving the improvement of classification accuracy (up to 96\%) when redshift information is included. In our subsequent analysis of the SN Ia rate, we used the results of \texttt{PSNID} with photometric redshift of the associated host galaxy as a prior. The only effect of mis-classified SNe is a decrease in the number of sources, because no other SN type is included in the simulation. Real life scenario would also include contamination from other transients, especially from SNe of type Ib/c. 

The SN Ia rate was measured on different samples to evaluate separately the impact of uncertainties due to photometric redshift, host galaxy association and classification. For each sample, we divided SN Ia in redshift bins of width 0.05 (which is the typical error of a photometric redshift estimate) in the range $0.1 \le z_{spec} \le 0.7$ and normalized the rate to the input model of the DC2 simulation. The different distribution in redshift of the various samples led to an average 17\% mismatch in the recovered fraction of SN Ia with respect to the original SN Ia sample. As 10\% of it is due to photometric redshifts alone (which also affect the classification), having good photometric redshift estimates results to be the major issue in the measurement of the SN Ia rate. The uncertainties not only change the number of SN Ia in each redshift bin, but also change the evolutionary trend of the recovered rate, hampering the discrimination between different progenitor models. 

Despite our estimate of the uncertainties might be reduced in the near future (e.g., using better algorithms, improving the measurement procedure, including additional information, etc.), we showed they are still relevant and have a significant impact on the rate measurement. Their reduction will be fundamental for precision SN Ia science, both for cosmology and stellar evolution studies. As the major source of uncertainty is due to photometric redshifts, improving their accuracy will be a priority. \cite{2020AJ....159..258G} demonstrated that adding near-infrared and near-ultraviolet photometry from the Euclid, Wide-Field InfrarRed Survey Telescope (WFIRST), and/or Cosmological Advanced Survey Telescope for Optical and ultraviolet Research (CASTOR) space telescopes can reduce both the standard deviation of photometric redshift estimates and the fraction of catastrophic outliers. The combination of Rubin and Euclid data would bring significant improvements also for other transient detection systematics, in particular for the estimate of the dust extinction bias (\citealt{2022zndo...5836022G}). A different scenario is expected for the DDF, where improvements in the photometric redshifts could certainly derive from the wealth of ancillary multi-wavelength data, with the drawback of using a smaller area and reducing the SN statistics. Spectroscopic follow up with other facilities could also provide a good sample of galaxies properties for SN rate analysis (albeit limited to low redshift). Finally, novel deep learning techniques to measure photometric redshift have shown to outperform other methods and might also be evaluated for LSST (e.g., \citealt{2019A&A...621A..26P}).

We used progenitor models from \cite{2005A&A...441.1055G} as a reference and adopted the SFH from \cite{2017ApJ...840...39M} to get predicted SN Ia rates for different progenitors. We found the combination of the uncertainties analyzed in this work to be as large as the discrepancy between the rate predictions from different progenitors. However, the scatter between the rate measurements is significantly lower than that between rate measurements obtained comparing multiple surveys, thus confirming the enormous capability of LSST. It is also worth noticing that up to $z \sim 1.0$ different models have similar outcomes and it would be necessary going to higher redshift to distinguish between them. An improvement of the redshift coverage could be possibly attained through the combination of Rubin and Roman data, which is expected to detect SN Ia up to $z \sim 3$, further increasing both the dimension and the variety of the sample (\citealt{roman_sin}; \citealt{roman}).
Moreover, our analysis on the effect of the uncertainties is also important to measure the SN Ia rate as a function of host galaxy intrinsic colors or specific SFR (which results to be promising in separating the predictions of different models as shown in \citealt{2019A&A...625A.113G}). Unfortunately, correlations between SN types and host galaxy properties were not included in this simulation, but their investigation is subject to subsequent analysis.


\begin{acknowledgements}
This paper has undergone internal review in the LSST Dark Energy Science Collaboration. We thank the internal reviewers Francisco Forster, Dominique Fouchez, and Christopher Frohmaier for their comments. We also deeply thank Richard Kessler for his help and advice throughout the work, especially on photometric classification and photometric redshifts. 

The DESC acknowledges ongoing support from the Institut National de 
Physique Nucl\'eaire et de Physique des Particules in France; the 
Science \& Technology Facilities Council in the United Kingdom; and the
Department of Energy, the National Science Foundation, and the LSST 
Corporation in the United States.  DESC uses resources of the IN2P3 
Computing Center (CC-IN2P3--Lyon/Villeurbanne - France) funded by the 
Centre National de la Recherche Scientifique; the National Energy 
Research Scientific Computing Center, a DOE Office of Science User 
Facility supported by the Office of Science of the U.S.\ Department of
Energy under Contract No.\ DE-AC02-05CH11231; STFC DiRAC HPC Facilities, 
funded by UK BEIS National E-infrastructure capital grants; and the UK 
particle physics grid, supported by the GridPP Collaboration.  This 
work was performed in part under DOE Contract DE-AC02-76SF00515.

V.P. and M.T.B. were responsible for the overall analysis and interpretation of the data, and wrote the paper. E.C. contributed to the analysis and interpretation of the results. L.G. provided theoretical models for SN Ia progenitor systems. B.O.S. provided data from his previous work and hints on the analysis. A.M. contributed to photometric classification with \texttt{SuperNNova}, while M.S. contributed to photometric classification with \texttt{PSNID}. M.L.G. provided comments on photometric redshifts and helped exploiting DP0 and RSP resources. M.P. and F.B. provided final comments on the analysis and the results. All authors contributed with notes and comments to improve the clarity of the paper.

A.M. is supported by the Australian Research Council Discovery Early Career Researcher Award (ARC DECRA) project number DE23010005. M.S. acknowledges support from DOE grant DE-FOA-0002424 and NSF grant AST-2108094. M.P. and V.P. acknowledge the financial contribution
from PRIN-MIUR 2022 funded by the European Union –
Next Generation EU, and from the Timedomes grant within the “INAF
2023 Finanziamento della Ricerca Fondamentale”.

The research has made use of the following \texttt{Python} software packages: \texttt{Astropy} (\citealp{Astropy2013,Astropy2018}), \texttt{Matplotlib} (\citealt{Hunter2007}), \texttt{Pandas} (\citealt{McKinney2010}), \texttt{NumPy} (\citealt{vanderwalt2011}), \texttt{SciPy} (\citealt{Virtanen2020}). Other software specific for SN analysis are cited in the paper.  

\end{acknowledgements}

%
%

\nocite{1999A&A...351..459C,2000A&A...362..419H,2002ApJ...577..120P,2003ApJ...594....1T,2004A&A...423..881B,2008A&A...479...49B,2008MNRAS.389.1871H,2008ApJ...681..462D,2010ApJ...713.1026D,2010ApJ...713.1026D,2011MNRAS.412.1441L,2010ApJ...723...47R,2011MNRAS.417..916G,2012A&A...545A..96M,2012ApJ...745...31B,2012AJ....144...59P,2013MNRAS.430.1746G,2014ApJ...783...28G,2014PASJ...66...49O,2014AJ....148...13R,2016yCat..35840062C,2019MNRAS.486.2308F}
\bibliography{references.bib}
\bibliographystyle{aa.bst}

\end{document}